\documentclass[final]{aipproc}

\usepackage{amsmath, amssymb, latexsym}

\layoutstyle{6x9}

\begin{document}

\title{When is Enough Good Enough\\in Gravitational Wave Source Modeling?}



\author{Louis J. Rubbo}{
  address={Center for Gravitational Wave Physics, 104 Davey Lab,
  University Park, PA 16802} }

%#### MAIN DOCUMENT ###############################

%==== Abstract ====================================

\begin{abstract}
A typical approach to developing an analysis algorithm for analyzing
gravitational wave data is to assume a particular waveform and use its
characteristics to formulate a detection criteria.  Once a detection
has been made, the algorithm uses those same characteristics to tease
out parameter estimates from a given data set.  While an obvious
starting point, such an approach is initiated by assuming a single,
correct model for the waveform regardless of the signal strength,
observation length, noise, etc.  This paper introduces the method of
Bayesian model selection as a way to select the most plausible
waveform model from a set of models given the data and prior
information.  The discussion is done in the scientific context for the
proposed Laser Interferometer Space Antenna.
\end{abstract}

\maketitle

%==== Main matter =================================

\section{INTRODUCTION} \label{sec:intro}

The anticipated data from the proposed Laser Interferometer Space
Antenna (LISA) introduces a number of exciting and original
challenges.  Central in these challenges is the development of data
analysis routines capable of coaxing out and characterizing individual
signals from the noisy time series LISA will return.  A great deal of
work has already been invested into the development of algorithms
applicable to the LISA data.  While a number of these algorithms have
demonstrated favorable capabilities on simulated data, each make an
initial assumption about the functional form for the waveform under
consideration.  This paper introduces the use of Bayesian model
selection as a quantitative method to selecting the waveform model.
Using Bayes' theorem we show how the data and prior information picks
out the most plausible model from a set of proposed models.

\begin{figure}
  \includegraphics[height=.27\textheight]{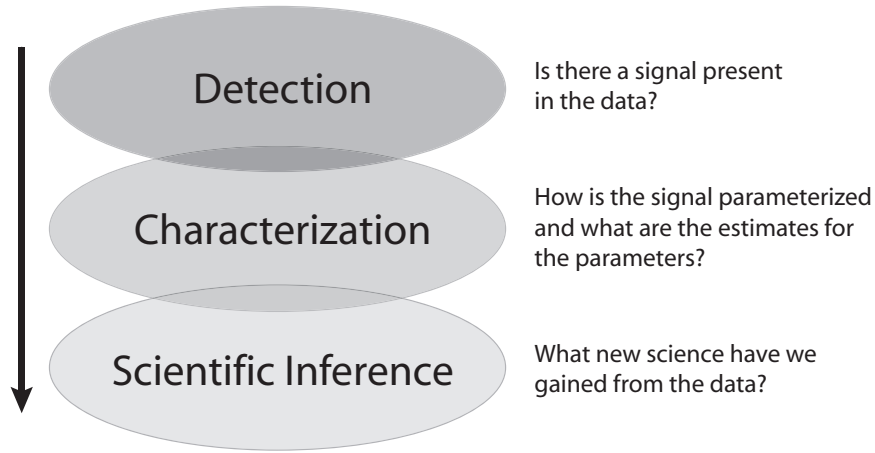}
  \caption{Data analysis flow chart.}
  \label{fig:flowchart}
\end{figure}

Gravitational wave data analysis can be loosely described as a three
step process as depicted in figure~\ref{fig:flowchart}.  In the first
step, a signal is detected within a set of noisy time streams
retrieved from the detector.  In step two, the signal is characterized
by producing estimates for the parameterization variables.  Finally,
step three is to make physical interpretations based on the estimated
parameter values.  These steps are not necessarily mutually exclusive.
There are no obvious boundaries and areas of overlap do exist.
However, each step is necessary when analyzing a detected signal.

In making the transition form detection to characterization (and quite
often in the detection process itself) a particular waveform is
assumed prior to the investigate.  While an obvious assumption to make
in the early developmental stages for an algorithm, it can lead to
needless complications and even misidentifications.  For example, if a
signal is characterized by a low signal-to-noise ratio, some of the
intricate waveform features can be lost in the noise and therefore a
simpler model would have sufficed in the analysis. In the Bayesian
model selection approach presented here, the data and prior
information justify the selection of a particular waveform model by
calculating the most plausible model from a proposed library of
models.

Bayesian model selection is not a new methodology, but it is one that
has not been fully adopted by the still infant gravitational wave
community.  The aim of this paper is to briefly summarize the theory
and to discuss possible applications for analyzing the LISA data.  To
this end, the paper first introduces the rules of probability theory,
including a derivation of Bayes' theorem.  It then outlines the
necessary calculations for performing a model selection procedure.
From here we give a simple, qualitative example of its use for the
LISA data.  We conclude by suggesting a few other applications
associated with LISA.

\section{BAYESIAN STATISTICS} \label{sec:bayes}

\subsection{Rules of Probability Theory} \label{sec:rules}

We begin by introducing a notation first used by
Jeffreys~\cite{Jeffreys:1961}.  We will denote the statement ``the
probability that proposition $A$ is true given proposition $B$'' as
$P(A|B)$.  Similarly, ``the joint probability that both $A$ and $B$
are true given $C$'' is denoted by $P(A,B|C)$.  The notation ``$|C)$''
is the conditional that proposition $C$ is assumed to be true.  In
Bayesian statistics probability statements such as $P(A)$ are not
clear because they do not explicitly state their dependencies.
Furthermore, \textit{all} probabilities are conditional.

Starting with the desiderata that degrees of plausibility are
represented by real numbers, the rules for manipulating plausibility
statements should agree with common sense, and they should be
consistent, then it is possible to show that the only two rules are
required for manipulating probabilities~\cite{Cox:1946}: the Sum Rule,
\begin{equation} \label{eq:sum_rule}
  P(A+B|C) = P(A|C) + P(B|C) - P(A, B|C)
\end{equation}
where the plus sign inside the probability argument means ``or'', and
the Product Rule,
\begin{equation} \label{eq:multi_rule}
  P(A,B|C) = P(A|C) P(B|A,C) \;.
\end{equation}
By standard Aristotelian logic it must be the case that $P(A,B|C) =
P(B,A|C)$.  Consequently, the Product Rule may be re-expressed as
\begin{equation}
  P(B,A|C) = P(B|C) P(A|B,C) \;.
\end{equation}
Equating the last two expressions results in Bayes' theorem,
\begin{equation} \label{eq:Bayes}
  P(A|B,C) = P(A|C) \; \frac{P(B|A,C)}{P(B|C)} \;.
\end{equation}
Although Bayes' theorem receives the accolades, it is simply a
consistency statement for the Product Rule.

In words, Bayes' theorem is often stated as
\begin{displaymath}
  \textrm{Posterior} = \textrm{Prior} \;
  \frac{\textrm{Marginal Likelihood}}{\textrm{Global Likelihood}} \;.
\end{displaymath}
In this form it is evident that Bayes' theorem quantitatively
describes a learning process.  We start with a prior state of
knowledge about proposition $A$ when $C$ is assumed true, $P(A|C)$.
We then gain new information $B$, which in return updates our final
state of knowledge as given by the posterior probability, $P(A|B,C)$.
The proportionality factor between our prior and posterior states of
knowledge is a normalized statement about how likely the proposition
$B$ will occur given that both $A$ and $C$ are true.

While Bayes' theorem is a useful byproduct of the Product Rule, the
use of the Sum Rule is equally important.  It is through the Sum Rule
that we are able to take a joint probability of multiple propositions,
and reduce it to a distribution of a smaller subset of the larger
joint distribution.  For example, consider the joint distribution
between $A$ and a set of $n$ exhaustive $B_{i}$'s, given prior
information $I$.  From the Sum Rule we have
\begin{eqnarray}
  P(A, \sum_{i=1}^{n} B_{i} | I) &=& P(A|I) \nonumber\\
  &=& P(A, B_{1} | I) + P(A, \sum_{i=2}^{n} B_{i} | I) - P(A, B_{1},
  \sum_{i=2}^{n} B_{i} | I) \;,
\end{eqnarray}
where the first equality follows from the Product Rule and the fact
that the $B_{i}$'s are exhaustive.  If the $B_{i}$'s are mutually
exclusive, that is only one value can be realized at a time, then the
last term is zero.  Repeated applications of the Sum Rule leads to 
\begin{equation}
  P(A|I) = \sum_{i=1}^{n} P(A, B_{i} | I) \;.
\end{equation}
When the $B_{i}$'s take on continuous values the above goes over to an
integral,
\begin{equation}
  P(A | I) = \int P(A, B | I) \, dB \;.
\end{equation}
The process which we have just described is referred to as
\textit{marginalization}.  In it we have removed a \textit{nuisance
parameter}, $B$, from a joint distribution by a repeated application
of the Sum Rule.

\subsection{Model Selection} \label{sec:ModelSel}

In model selection the central question that is being addressed is the
following: ``Given a particular set of data, and prior information,
which hypothesis from a library $\mathcal{L} \equiv \{H_{1}, \ldots,
H_{\ell}\}$ of hypotheses is the most plausible?''  Key to this
question are the ideas that all prior information is included and that
the most plausible hypothesis is based on the given data.  The
hypotheses within a library are either assumed to be exhaustive or, by
a careful choice in models, the space is made
so~\cite{Bretthorst:1996}.

A model itself consists of a functional form dependent on a vector of
parameters $\vec{\lambda}$, and two probability
distributions~\cite{MacKay:1992}.  The first distribution describes
the probability distribution for the parameter values given the model
prior to the new data, $P(\vec{\lambda}|H_{\alpha})$.  This is a key
point; two models are distinct even if they have the same
parameterization but different priors about how those parameters are
believed to be distributed.  The second distribution is the
probability of a data set given the model and a particular set of
parameter values, $P(D|\vec{\lambda}, H_{\alpha})$.

From Bayes' theorem~\eqref{eq:Bayes}, the posterior probability for a
particular model is given by
\begin{equation}
  P(H_{\alpha}|D, I) = P(H_{\alpha}|I) \; \frac{P(D|H_{\alpha}, I)}{
  P(D| I)} \;, 
\end{equation}
where $I$ symbolizes our unenumerated prior information.  The
denominator can be viewed as a normalization constant,
\begin{equation}
  P(D|I) = \sum_{\alpha = 1}^{\ell} P(H_{\alpha}|I) P(D|H_{\alpha}, I)
  \;.
\end{equation}
By investigating the \textit{odds ratio} between two competing models,
we can eliminate the need to calculate the normalization constant,
\begin{eqnarray} \label{eq:oddsratio}
  O_{12} &=& \frac{P(H_{1}|D,I)}{P(H_{2}|D,I)} =
  \frac{P(H_{1}|I) P(D|H_{1},I)}{P(H_{2}|I) P(D|H_{2},I)} \nonumber\\
  &=& \frac{P(D|H_{1},I)}{P(D|H_{2},I)} \;.
\end{eqnarray}
The second line arises by assuming that our prior information does not
favor one model over the other.  The odds ratio gives us a means to
directly compare competing models.  If our library contains more than
two models, one model may be used as a reference.  For example, the
reference model may be a constant (i.e. a no signal present model),
while the remaining library contains a spectrum of waveform models.

From the odds ratio it is apparent that to compare models in a library
only their marginal likelihoods need to be calculated.  The
likelihoods are found by marginalizing, over all model parameters, the
joint distribution for the data and the model parameters,
\begin{equation} \label{eq:evidence}
  P(D|H_{\alpha}, I) = \int P(D, \vec{\lambda}_{\alpha} | I) \;
  d\vec{\lambda}_{\alpha} = \int P(\vec{\lambda}_{\alpha}|H_{\alpha},
  I) P(D|\vec{\lambda}_{\alpha}, H_{\alpha}, I) \;
  d\vec{\lambda}_{\alpha} \;,
\end{equation}
where the second equality follows from the Product Rule.

\begin{figure}
  \includegraphics[height=0.27\textheight]{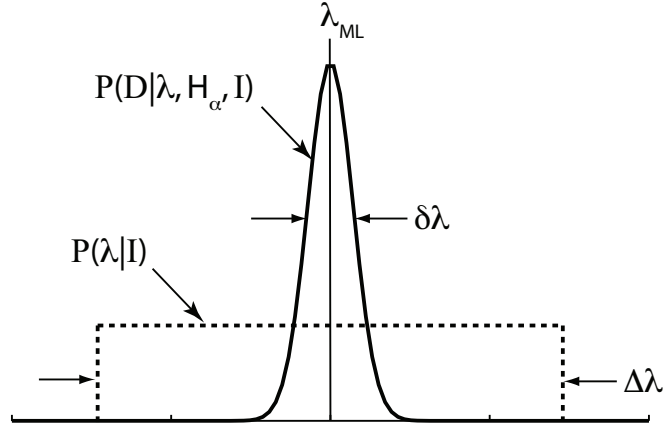}
  \caption{A pictorial representation for the origins of Occam factors
    in Bayesian model comparisons.}
  \label{fig:occam}
\end{figure}

If the data is informative, i.e. we have learned something new, then
the parameter likelihood function, $P(D|\vec{\lambda}_{\alpha},
H_{\alpha}, I)$, will be more peaked than the parameter priors,
$P(\vec{\lambda}_{\alpha} | H_{\alpha}, I)$.  Figure~\ref{fig:occam}
illustrates this for a one dimensional model.
In this instance we can estimate the marginal likelihood as
\begin{equation}
  P(D|H_{\alpha},I) \approx P(D | \lambda_{ML}, H_{\alpha}, I) \left[
  P(\lambda_{ML}|H_{\alpha}, I) \; \delta\lambda \right] \;.
\end{equation}
Here $\lambda_{ML}$ is the parameter value at the maximum likelihood
and $\delta\lambda$ is the characteristic width for the parameter
likelihood function.  The term in square brackets is an \textit{Occam
factor}; a term that naturally penalizes complicated models. To see
this consider a uniform prior, $P(\lambda|I) = (\Delta\lambda)^{-1}$,
where $\Delta\lambda$ is the interval width for the range of expected
parameter values before the data is collected.  The marginal
likelihood is now
\begin{equation}
  P(D|H_{\alpha},I) \approx P(D | \lambda_{ML}, H_{\alpha}, I) \;
  \frac{\delta\lambda}{\Delta\lambda} \;.
\end{equation}
For informative data the Occam factor is always less than unity.
Consequently, for a complicated model to be favored over a simpler
one, the data must justify it by having a corresponding larger value
for the parameter likelihood function.

The proceeding argument is quickly extended to multiple dimensions.
If the model has more than one parameter, then there is a
corresponding Occam factor for each parameter,
\begin{equation} \label{eq:apprxpost}
  P(D|H_{\alpha},I) \approx P(D | \vec{\lambda}_{ML}, H_{\alpha}, I)
  \; \frac{\delta\lambda_{1}}{\Delta\lambda_{1}} \cdots
  \frac{\delta\lambda_{i}}{\Delta\lambda_{i}} \;,
\end{equation}
where $i$ is the number of parameters.

As a last point of emphasis, it is not enough to perform a parameter
estimation analysis and find that $\lambda_{i} = 0$, therefore ruling
out the model that includes $\lambda_{i}$.  Doing so would neglect the
Occam factors that arise in Bayesian model selection and are not
present in a parameter estimation analysis, even a Bayesian analysis.

\section{WHITE DWARF TRANSFORM} \label{sec:wdtrans}

As a conceptually trivial but applicable example of Bayesian model
selection for the LISA mission, consider the detection of a
supermassive black hole binary inspiral.  For black hole binaries with
component masses in the range of $10^{4-7}~\textrm{M}_{\odot}$, LISA
will observe the binary evolution as the binary sweeps through
frequencies from $\sim\!0.01$~mHz up to a few milliHertz (depending on
the actual masses).  In this same range of frequencies is the
gravitational wave background formed from the $\sim\!10^{8}$ solar
mass binaries in our own galaxy.  As the black holes inspiral, their
detected signal will overlap with the collective galactic background
signal.  Moreover, at any instant of time the black hole binary looks
like a monochromatic binary.  That is, as a supermassive black hole
binary with a time to coalescence of $t_{c}$ sweeps past a galactic
binary of period $T$, the two signals have a significant overlap for
an interval equal to the geometric mean of $t_{c}$ and
$T$~\cite{Cornish:2005}.  Consequently the black hole inspiral signal
may be decomposed into a population of monochromatic galactic
binaries.  Such a process is often referred to as a \textit{white
dwarf transform}.

For a gravitational wave data analyst the task is to select which of
two models is more plausible.  The models under consideration are
\begin{eqnarray*}
  H_{WD} &=& \left( \begin{array}{l} \text{the detected signal is from
  a population} \\ \text{of monochromatic galactic binaries}
  \end{array} \right) \\
  H_{BH} &=& \left( \begin{array}{l} \text{the detected signal is from
  a single} \\ \text{supermassive black hole binary} \end{array}
  \right) \;.
\end{eqnarray*}
Model $H_{WD}$ is parameterized by $7N$ variables, where $N$ is the
number of binaries required to describe the apparent inspiral signal.
For an inspiral signal between $0.01$ and 1~mHz, $N$ is on the order
of $10^{4}$ assuming a binary per frequency bin and for a one year
observation\footnote{A frequency bin $\Delta f$ is equal to one on the
observation time, $\Delta f = T^{-1}$.  For a one year observation,
which is used here, $\Delta f = 3.2 \times 10^{-8}$~Hz.}.  Conversely,
model $H_{BH}$ is characterized by only seventeen parameters.

Estimating the posterior probabilities using
equation~\eqref{eq:apprxpost} quickly leads to the conclusion that the
large parameter space associated with the white dwarf population model
has associated with it an overwhelming number of Occam factors.  These
Occam factors penalize the white dwarf population model and in turn
make the plausibility for the model extremely low.  The black hole
model, on the other hand, only has seventeen Occam factors and
therefore is not as severely penalized.  Consequently, although an
ensemble of galactic binaries could conspire to look like a
supermassive black hole binary inspiral, the relative probability for
such a model is many orders of magnitude less than a model that
contains a single black hole binary.

\section{CONCLUDING REMARKS} \label{sec:conclusions}

The white dwarf transform is an obvious application of Bayesian model
selection.  More informative and interesting examples include using
Bayesian model selection as a criteria for deciding when a signal is
present in the data; characterizing complicated but detected signals
that have low signal-to-noise ratios; and counting the number of
detectable galactic binaries within the larger population.  The first
application is simply answering the question, when does the data
justify declaring a detection for a particular waveform?  The second
application is concerned with deciding the information content from a
weak signal.  That is, what features of an emitting system are
actually measurable and what features are lost to the noise.  Counting
the number of detectable galactic binaries is one of the few Bayesian
model selection applications used in the LISA
literature~\cite{Umstatter:2005a, Stroeer:2006}.  Embedded within
Reversible Jump Markov Chain Monte Carlo techniques is the use of odds
ratios in deciding the number of galactic binaries that are
detectable.

In general, Bayesian model selection gives a logical and quantitative
approach to directly comparing competing models.  By using a model
selection procedure we are able to maximize the amount of information
we can extract from LISA's data.  The most plausible model is the one
that is most justified by the data and our prior state of knowledge
prior to the experiment.  As progress is made in the development of
LISA analysis routines it is conceivable that Bayesian approaches will
be a central tool.

%==== Acknowledgments =============================

\begin{theacknowledgments}
The author would like to thank Edward Cazalas, Matthew Francis, and
Deirdre Shoemaker for a number of helpful discussions.  Also, Lee
Samuel Finn for introducing the author to the Bayesian approach and
for guidance on a number of its subtler points.  This work was
supported by the Center for Gravitational Wave Physics.  The Center
for Gravitational Wave Physics is funded by the National Science
Foundation under cooperative agreement PHY 01-14375.
\end{theacknowledgments}

%==== Bibliography ================================

\bibliographystyle{aipproc}
\bibliography{References}

\begin{thebibliography}{7}
\expandafter\ifx\csname natexlab\endcsname\relax\def\natexlab#1{#1}\fi
\providecommand{\enquote}[1]{``#1''}
\expandafter\ifx\csname url\endcsname\relax
  \def\url#1{\texttt{#1}}\fi
\expandafter\ifx\csname urlprefix\endcsname\relax\def\urlprefix{URL }\fi
\providecommand{\eprint}[2][]{\url{#2}}

\bibitem[Jeffreys(1961)]{Jeffreys:1961}
H.~Jeffreys, \emph{{Theory of Probability}}, The International Series of
  Monographs on Physics, Clarendon Press, Oxford, 1961, 3 edn.

\bibitem[Cox(1946)]{Cox:1946}
R.~T. Cox, \emph{American Journal of Physics} \textbf{14}, 1--13 (1946).

\bibitem[Bretthorst(1996)]{Bretthorst:1996}
G.~L. Bretthorst, \enquote{{An Introduction to Model Selection using
  Probability Theory as Logic},} in \emph{Maximum Entropy and Bayesian
  Methods}, edited by G.~R. Heidbreder, Kluwer Academic Publishers, Dordrecht,
  1996, pp. 1--42.

\bibitem[MacKay(1992)]{MacKay:1992}
D.~J.~C. MacKay, \emph{Neural Systems} \textbf{4}, 415--447 (1992).

\bibitem[Cornish and Crowder(2005)]{Cornish:2005}
N.~J. Cornish, and J.~Crowder, \emph{Physical Review D} \textbf{72}, 043005
  (2005).

\bibitem[Umst{\"a}tter et~al.(2005)]{Umstatter:2005a}
R.~Umst{\"a}tter, et~al., \emph{Physical Review D} \textbf{72}, 022001 (2005).

\bibitem[Stroeer and Vecchio(2006)]{Stroeer:2006}
A.~Stroeer, and A.~Vecchio, {Automatic Bayesian Inference for LISA Data
  Analysis} (2006), presentation at the Sixth International LISA Symposium.

\end{thebibliography}

\end{document}